%% file: newmain.tex
\documentclass[a4paper, 11pt]{article}
\input{bib_macros}

\usepackage{geometry}
\usepackage[numbers]{natbib}
\usepackage[utf8]{inputenc}
\usepackage{enumitem,amssymb}
\usepackage{ragged2e}
\usepackage{setspace}
\usepackage{eso-pic,graphicx}
\usepackage{tikz}
\usepackage{subfigure}

\usepackage{xcolor}
\definecolor{cobalt}{rgb}{0., 0.35, 0.56}
\usepackage{hyperref}
\hypersetup{colorlinks=true,
            citecolor=cobalt,
            linkcolor=cobalt,
            urlcolor=cobalt,
            pdftitle={A new path to constrain the expansion history of the Universe in future spectroscopic galaxy surveys}}

\begin{document}

\raggedright
\Large
ESO Expanding Horizons initiative 2025 \linebreak
Call for White Papers

\vspace{2.cm}
\begin{spacing}{1.6}
\textbf{\fontsize{22pt}{40pt}\selectfont
A new path to constrain the expansion history of the Universe in future spectroscopic galaxy surveys}
\end{spacing}
\normalsize

\vspace{0.5cm}

\textbf{Authors:} Elena Tomasetti$^\mathbf{1,2}$, Michele Moresco$^\mathbf{1,2}$, Nicola Borghi$^\mathbf{1,2,6}$, Dinko Milakovi{\'c}$^\mathbf{3,4}$, Stephanie Escoffier$^\mathbf{5}$, Margherita Talia$^\mathbf{1,2}$, Lucia Pozzetti$^\mathbf{2}$, Andrea Cimatti$^\mathbf{1,2}$, Lauro Moscardini$^\mathbf{1}$

\vspace{0.3cm}
\textbf{Contact:} \href{mailto:elena.tomasetti2@unibo.it}{elena.tomasetti2@unibo.it}
\linebreak

\textbf{Affiliations:} \\
{\footnotesize
$^\mathbf{1}$ Dipartimento di Fisica e Astronomia ``Augusto Righi'' -- Universit\`{a} di Bologna, via Gobetti 93/2, I-40129 Bologna, Italy \\
$^\mathbf{2}$ INAF-OAS, Osservatorio di Astrofisica e Scienza dello Spazio di Bologna, via Gobetti 93/3, I-40129 Bologna, Italy \\
$^\mathbf{3}$ INAF-OATs, Osservatorio di Trieste, Via Tiepolo 11, I-34131 Trieste, Italy \\
$^\mathbf{4}$ Institute for Fundamental Physics of the Universe (IFPU), via Beirut 2, I-34151 Trieste, Italy \\
$^\mathbf{5}$ Aix Marseille Université, CNRS/IN2P3, CPPM, Marseille, France\\
$^\mathbf{6}$ INFN - Sezione di Bologna, Viale Berti Pichat 6/2, I-40127 Bologna, Italy\\
}

\pagenumbering{gobble} 

\pagebreak

\justifying

\section*{Opening a new avenue to constrain the expansion rate of the Universe}

The measurement of the Hubble constant ($H_0$) is one of the most fundamental yet debated aspects of modern cosmology. Discrepancies in the measurements of $H_0$ from the early Universe (e.g., CMB) and late Universe (e.g., SNIa, BAO) have led to the Hubble tension \citep{kamionkowski_hubble_2023}. Additionally, the nature of dark matter and dark energy and their role in the acceleration of the Universe’s expansion remain far from being fully understood. In this context, deriving a direct mapping with minimal cosmological assumptions of the expansion history of the Universe through the Hubble parameter, $H(z)$, is crucial. Such measurements provide a powerful and complementary means of constraining cosmological parameters independently of standard probes and, most importantly, in a model-independent way.

The cosmic chronometers \citep[CC,][]{jimenez_constraining_2002} approach offers the unique opportunity to reconstruct the evolution of the Hubble parameter in redshift mapping the ageing of a carefully selected population of very massive and passively evolving galaxies, applying the equation: $H(z) = \frac{-1}{1+z}\frac{\rm{d}z}{\rm{d}t}$.
Since this equation is obtained without assuming any cosmological model, the $H(z)$ reconstruction that can be obtained through this method represents a perfect test bench for alternative cosmological models and for constraining cosmological parameters and their evolution. To ensure the homogeneity of the sample, though, a detailed process of selection is required, dramatically reducing the sample size. On the other hand, the numerosity of the sample mapping the age-redshift is fundamental to reconstruct its slope (d$z$/d$t$). For these reasons, the statistics of the survey is crucial in the CC context.

Currently, the CC method has effectively mapped the expansion history of the Universe up to redshift z$\lesssim$0.6, primarily through the Sloan Digital Sky Survey (SDSS) \citep[see, e.g.,][]{moresco_new_2012,moresco_6_2016}. However, at higher redshifts, the absence of a comprehensive, homogeneous, and statistically significant survey results in a sparse $H(z)$ mapping, with typical uncertainties exceeding 20\%. In this context, a homogeneous spectroscopic survey with high statistics, wide redshift coverage, and high-quality spectra in terms of resolution and signal-to-noise ratio ($S/N$) is essential to fully unlock the potential of this probe.

\section*{State of the art and current limitations}

The application of the CC method requires a rigorous selection process to isolate a pure sample of massive, passively evolving galaxies, ensuring maximum homogeneity of the sample, whose ageing can probe the ageing of the Universe in redshift. This selection typically proceeds through multiple stages: initial photometric cuts to identify the reddest galaxies, followed by spectroscopic criteria to confirm the absence of emission lines and ongoing star formation, and finally the use of spectroscopic diagnostics to remove objects with residual contamination by younger components \citep[see][]{moresco_unveiling_2022}. Each of these steps progressively reduces the sample size to ensure the purity required for reliable age-dating, with up to 80--85\% of the initial sample being discarded \citep{tomasetti_new_2023,borghi_toward_2022}.

To date, no spectroscopic survey has been specifically designed or optimised for CC studies. As a consequence, all CC measurements to date rely on legacy data from surveys conceived for different scientific goals, with selection criteria and observational strategies not tailored to the specific needs of the CC method. This has resulted in heterogeneous samples, limited statistics, and, in many cases, S/N, and incomplete redshift coverage.

Nevertheless, various studies have demonstrated that competitive constraints on $H(z)$ can be achieved when sufficiently large datasets are available \citep[see, e.g.,][]{moresco_improved_2012,moresco_6_2016,jimenez_cosmic_2023}. A notable example is the work of \citet{moresco_6_2016}, which exploited the SDSS-BOSS survey to assemble a sample of $\sim$130,000 CCs. From this dataset, five independent $H(z)$ measurements were derived in the range $0.38 < z < 0.47$, each reaching a precision of 4--5\%. When combined, these yielded a single constraint with a remarkable precision of 1.1\% (stat). This clearly illustrates the potential of the CC method when high statistics are available, and highlights how a future survey specifically designed for CC studies could push these constraints to unparalleled levels across a much broader redshift range. However, in the landscape up to 2040, no spectroscopic survey presents the requirements needed in terms of wavelength coverage, resolution, S/N ratio, and statistics at the same time.

\section*{Requirements and future prospects}

A next-generation wide-field spectroscopic facility with a very large multiplex capability can overcome this limitation ($O(10^4)$). Targeting massive and passive galaxies to a depth of $z \sim 22 -23$ mag would provide access to number densities of $\sim$2000--3000  deg$^{-2}$ luminous red galaxies (LRGs), mapping the massive end of the galaxy population up to $z \sim 1.2$, far beyond the redshift covered by other samples. To fully exploit such a dataset, a $S/N \geq 10$ \AA$^{-1}$ is essential for reliable age-dating via full spectral fitting \citep{citro_methodology_2017,carnall_vandels_2019}, while a spectral resolution of $R \gtrsim 3000$ ensures that key diagnostic features, like the 4000 \AA\ break and metal absorption lines, are adequately resolved. A high multiplexing power is required not only to efficiently survey the large number of targets needed, but also to provide a sufficiently dense sampling to allow the characterization of the population, also as a function of the environment.

As an example, the Wide-field Spectroscopic Telescope \citep[WST,][]{bacon_wst_2024} could meet these requirements. With its advantageous combination of wide field of view, high multiplex capability, and sensitivity, WST could observe hundreds of millions of galaxies across a broad redshift range, including a substantial population of luminous red galaxies (LRGs) ideal for CC studies.
With tens of millions of passive galaxies in the range $0.3 < z < 1.2$ expected within the first years of operation, its statistical power would vastly surpass current data at lower redshifts. Such a sample would be transformative from different perspectives. Even under the conservative assumption that only a few percent ($\sim$10 million objects) of the total galaxy sample consists of passive galaxies, and that 80\% of these are discarded during the stringent selection process required for CC studies, the resulting sample would still comprise $\sim$2 million CCs spanning the range $0.3 < z < 1.2$. This is fifteen times the statistical power of the CC sample in SDSS-BOSS, which provided a single 1\%-precision $H(z)$ measurement. Such an improvement would enable, for the first time, a precise and continuous cosmology-independent mapping of the expansion history of the Universe out to $z \sim 1.2$, covering more than half of cosmic time with percent-level accuracy, and transforming the CC method into a fully competitive precision cosmological probe. 

\section*{Cosmological implications}

\begin{figure*}[t]
    \centering
    \subfigure[]{
        \includegraphics[width=0.57\textwidth]{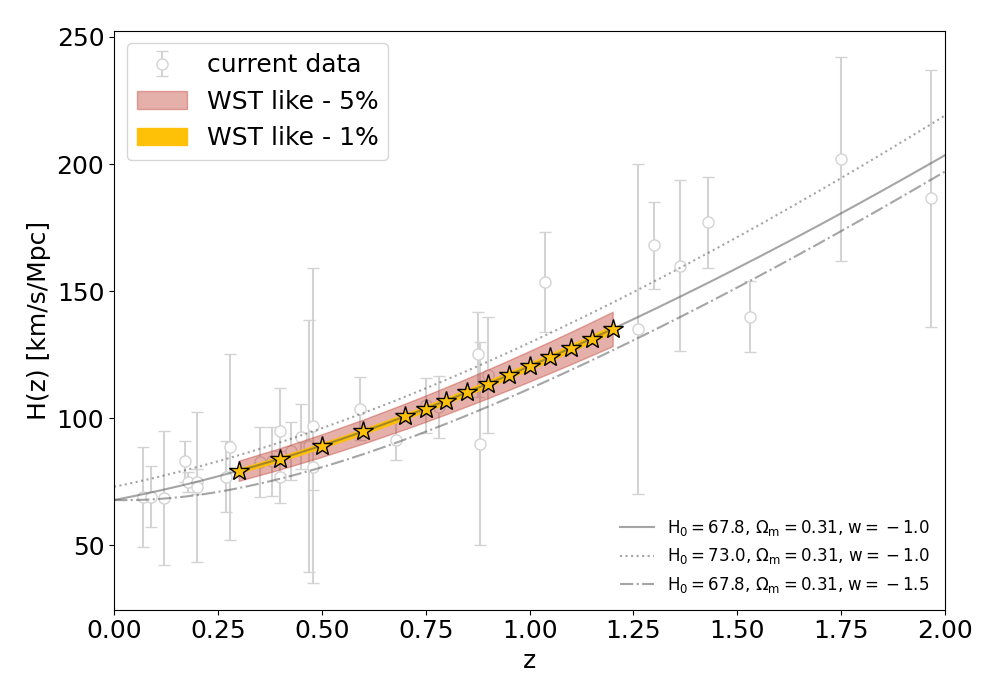}
    }
    \subfigure[]{
        \includegraphics[width=0.39\textwidth]{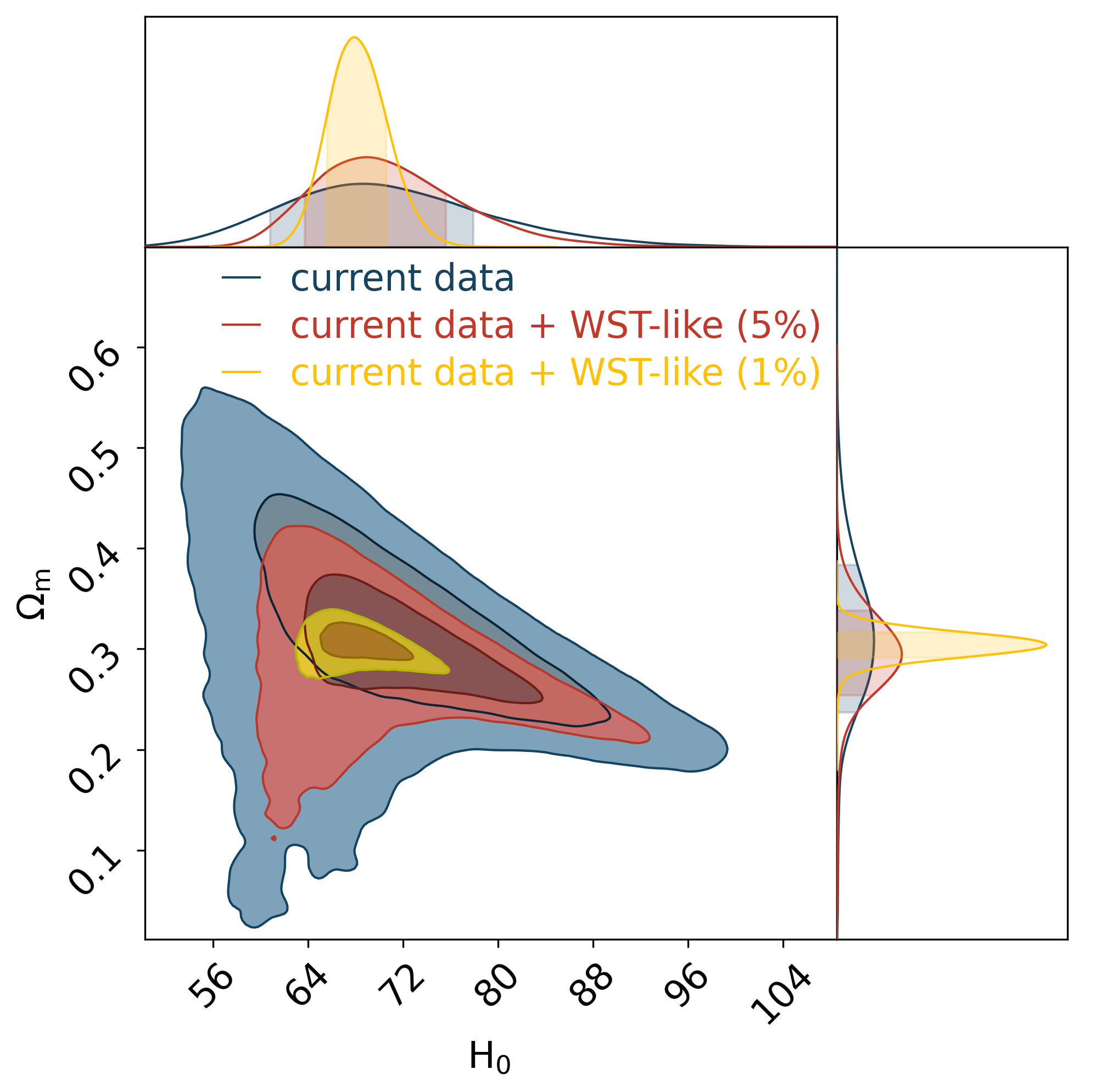}
    }
    \caption{\textit{Left:} Forecasts on the potential $H(z)$ measurements attainable with a WST-like facility. The shaded areas illustrate the 5\% (red) or 1\% (yellow) precision on the $H(z)$ points. In grey, the collection of $H(z)$ measurements obtained to date with the CC method. \textit{Right:} Constraints attainable in the $\Omega_m-H_0$ plane with different precisions on the CC dataset. The fit is performed within a flat $w$CDM framework. Corresponding results are in Table \ref{tab:model_params_opt}.}
    \label{fig:cosmo_fit}
\end{figure*}

To assess the potential of a future ESO facility for CC studies, we generate simulated $H(z)$ measurements representative of the expected performance of such a survey, as presented in Fig.~\ref{fig:cosmo_fit}. Based on the aforementioned estimates, we consider 15 independent $H(z)$ data points spanning $0.3 < z < 1.2$, with a denser sampling in the region $0.8 < z < 1.2$. The mock data are generated assuming fiducial cosmological parameters from \citet{collaboration_planck_2020} in a flat $w$CDM, exploring two scenarios: a conservative configuration with 5\% precision, and an optimistic one with 1\% precision, reflecting the expected performance based on the statistical power of the anticipated CC sample.

Following the approach used in \citep{moresco_unveiling_2022}, we estimate the error budget considering in the covariance matrix the different contributions that might affect this method \citep[see][for details]{moresco_unveiling_2022}. As in \cite{moresco2024}, we consider two scenarios: a standard one, where all systematic contributions are included, and an optimistic one, where some of the systematics uncertainties are assumed to be resolved. This second scenario reflects the expected improvements from ongoing efforts to calibrate stellar population models and reduce model-dependent systematics. Here, we provide forecasts that can be obtained in a flat $w$CDM cosmology, deriving constraints on the Hubble constant $H_0$, the matter density parameter $\Omega_m$, and the dark energy equation of state $w$.
Results are presented in Fig.~\ref{fig:cosmo_fit} and Tab.~\ref{tab:model_params_opt}. It is clear how the addition of WST-like measurements significantly improves these constraints: the optimistic scenario with 1\% precision reduces the uncertainty on $H_0$ to 3.7\% and on $w$ to 14.9\%. Remarkably, this level of precision on the dark energy equation of state is comparable to current constraints from DESI BAO alone \citep{adame_desi_2025}. It is important to emphasize that these forecasts are based exclusively on CCs, without combining with any other cosmological dataset. Further improvements are expected both from extending the CC sample to $z > 1.2$ with upcoming surveys \citep[e.g., MOONS,][]{cirasuolo_moons_2020}, and from the combination with other late-Universe probes \citep[e.g., time delay cosmography, see][]{bergamini_augmenting_2024}. The synergy between WST and these complementary datasets has the potential to deliver sub-percent constraints on $H_0$, together with an accurate reconstruction of the expansion history of the Universe up to high$-z$, opening a new precision era for late-Universe cosmology. Nevertheless, such a sample could represent a legacy treasure also for other applications, including multi-messenger gravitational wave cosmology with standard sirens, as well as comprehensive studies of galaxy formation and evolution.

\begin{table*}[t]
    \centering
    \caption{Cosmological constraints for a flat $w$CDM model from current CC data and forecasted WST-like measurements. The upper section shows results with the full systematic error budget, while the lower section presents an optimistic scenario where SPS model uncertainties are assumed to be resolved.}
    \label{tab:model_params_opt}
    \vspace{2mm}
    \begin{tabular}{lcccccc}
        \hline
        Dataset & $H_0$ & $\sigma_{\%}$ & $\Omega_m$ & $\sigma_{\%}$ & $w_{\rm DE}$ & $\sigma_{\%}$ \\ 
        \hline
        current data & $68.77^{+9.14}_{-7.98}$ & 12.4\% & $0.31^{+0.08}_{-0.07}$ & 23.8\% & $-1.17^{+0.56}_{-0.89}$ & 62.1\% \\ 
        current data + WST 5\% & $68.84^{+6.93}_{-6.59}$ & 9.8\% & $0.30^{+0.06}_{-0.05}$ & 18.2\% & $-1.20^{+0.46}_{-0.59}$ & 43.7\% \\ 
        current data + WST 1\% & $68.50^{+2.86}_{-2.88}$ & 4.2\% & $0.31^{+0.02}_{-0.03}$ & 8.0\% & $-1.08^{+0.26}_{-0.34}$ & 28.0\% \\
        \hline
        \multicolumn{7}{c}{Optimistic scenario} \\
        \hline
        current data + WST 5\% & $68.90^{+6.71}_{-5.18}$ & 8.6\% & $0.29^{+0.04}_{-0.04}$ & 14.4\% & $-1.14^{+0.39}_{-0.45}$ & 36.8\% \\ 
        current data + WST 1\% & $67.94^{+2.63}_{-2.35}$ & 3.7\% & $0.30^{+0.01}_{-0.01}$ & 4.2\% & $-1.03^{+0.16}_{-0.15}$ & 14.9\% \\ 
        \hline
    \end{tabular}
\end{table*}

\vfill
\bibliographystyle{aa_1auth}
\bibliography{references}

\end{document}

%% file: bib_macros.tex
\usepackage{paralist}

